\documentstyle[12pt]{article} 
\makeatletter \def\section{\@startsection {section}{1}{\z@}{-3.5ex
plus -1ex minus -.2ex}{2.3ex plus .2ex}{\large\bf}}
\def\subsection{\@startsection{subsection}{2}{\z@}{-3.25ex plus -1ex
minus -.2ex}{1.5ex plus .2ex}{\normalsize\bf}} \makeatother

\font\cmss=cmss10 \font\cmsss=cmss10 at 7pt \def\inbar{\vrule
height1.5ex width.4pt depth0pt}
\def\IC{\relax\,\hbox{$\inbar\kern-.3em{\rm C}$}}
\def\IG{\relax\,\hbox{$\inbar\kern-.3em{\rm G}$}} \def\IB{\relax{\rm
I\kern-.18em B}} \def\ID{\relax{\rm I\kern-.18em D}}
\def\IL{\relax{\rm I\kern-.18em L}} \def\IF{\relax{\rm I\kern-.18em
F}} \def\IH{\relax{\rm I\kern-.18em H}} \def\II{\relax{\rm
I\kern-.17em I}} \def\IN{\relax{\rm I\kern-.18em N}}
\def\IP{\relax{\rm I\kern-.18em P}}
\def\IQ{\relax\,\hbox{$\inbar\kern-.3em{\rm Q}$}}
\def\bfzero{\relax\,\hbox{$\inbar\kern-.3em{\rm 0}$}}
\def\IK{\relax{\rm I\kern-.18em K}}
\def\IG{\relax\,\hbox{$\inbar\kern-.3em{\rm G}$}} \font\cmss=cmss10
\font\cmsss=cmss10 at 7pt \def\IR{\relax{\rm I\kern-.18em R}}
\def\ZZ{\relax\ifmmode\mathchoice {\hbox{\cmss Z\kern-.4em
Z}}{\hbox{\cmss Z\kern-.4em Z}} {\lower.9pt\hbox{\cmsss Z\kern-.4em
Z}} {\lower1.2pt\hbox{\cmsss Z\kern-.4em Z}}\else{\cmss Z\kern-.4em
Z}\fi} \def\bfone{\relax{\rm 1\kern-.35em 1}}   
  
 \def\bar{\overline} 

\def\IE{\relax{{\rm I\kern-.18em E}}} 
 \def\IGam{\relax{{\rm I}\kern-.18em \Gamma}}
 
\def\beq{\begin{equation}} \def\eeq{\end{equation}}
\def\beqa{\begin{eqnarray}} \def\eeqa{\end{eqnarray}}
\def\nn{\nonumber}

\begin{document}

\begin{titlepage}
\setcounter{page}{0}

\begin{flushright}

NORDITA--2000/59 HE \\ SWAT/265

\end{flushright}

\vskip 25pt

\begin{center}

{\Large \bf Microscopic entropy of the most general 
four-dimensional BPS black hole}

\vskip 30pt

{\large  Matteo Bertolini}
\vskip 5pt {\it NORDITA \\ Blegdamsvej 17, DK-2100 Copenhagen \O,
Denmark} \\ {\footnotesize{teobert@nordita.dk, teobert@nbi.dk}}
\vskip 8pt
\vskip 8pt {\large  Mario Trigiante}
\vskip 5pt {\it Department of Physics, University of Wales Swansea \\
Singleton Park, Swansea SA2 8PP, United Kingdom} \\
{\footnotesize{m.trigiante@swan.ac.uk}}

\end{center}

\vskip 40pt

\begin{abstract}
In a recent paper we have given the macroscopic and microscopic
description  of the generating solution of toroidally compactified
string theory BPS  black holes. In this note we compute its
corresponding microscopic entropy.  Since by definition the generating
solution is the  most general solution modulo $U-$duality
transformations, this result allows  for a description of the
fundamental degrees of freedom accounting for the  entropy of any
regular BPS black holes of toroidally compactified string  (or M)
theory.
\end{abstract}

\end{titlepage}

\section{Introduction}
Recently we have given a description, both at macroscopic and
microscopic level, of the generating  solution of four dimensional
regular BPS black holes within toroidally compactified string (or M)
theory, \cite{gen}.  Acting by means of  $U$--duality transformations
on the above solution one can reconstruct any other  solution of the
same kind within the relevant four dimensional effective theory
(i.e. $N=8$ supergravity) and also give  for each of them a
corresponding microscopic description in terms of bound states of
stringy objects, \cite{mm}.  Because of its very definition, the
generating solution encodes the fundamental degrees of freedom
(related to $U$--duality invariants) characterizing {\it the most
general} regular BPS black hole within the same theory. Hence a
detailed microscopic understanding of this solution is of considerable
relevance for a deeper understanding  of stringy oriented microscopic
entropy counting\footnote{For a review see for example
\cite{malda,amanda} and, more recently, \cite{mooa} and references
therein.}.

Although we have given a prediction for the microscopic entropy of the
generating solution which is consistent with what  is expected (see
\cite{bala}), a missing aspect in our analysis was the {\it
statistical} interpretation (and computation)  of the predicted
formula. In this note we fill such a lack. This can be done by
suitably extending the microscopic computation of \cite{msw} (which
was carried out in the context of  Calabi-Yau compactifications) to
the toroidal case. One of the main concerns in the analysis of
\cite{msw} was to compute higher order corrections to the
semi--classical result. In general one expects both $\alpha'$ and
$g_{string}$ corrections. It is known (see \cite{msw,v1,moooa,sen1})
that the two affect the area law by a deformation  of the effective
horizon area and by a deviation from the area law itself. The latter
has a leading term which is topological (originating from $R^4$ terms
in M--theory) and which thus depends on the particular compactifying
manifold considered, giving therefore  different contributions for
Calabi--Yau and toroidal compactifications. Although our analysis is limited 
just to  the semiclassical
leading approximation we shall easily see that in  the case of tori
this topological term does not contribute.

The microscopic configuration corresponding to the generating solution
presented in \cite{gen} can be described, in the  type IIA framework,
as a bound state of 3 bunches of D4--branes intersecting on a point, a
bunch  of D0--branes on top of them plus some magnetic flux on the
D4--branes world volume (in a way which is consistent  with the
residual supersymmetry of the solution)  which induces D2 and (extra)
D0--brane effective charges. Upon $T$--duality transformations  along
three internal directions the same solution is described in type IIB
by four bunches of D3--branes intersecting at non--trivial $SU(3)$
angles (where the non--trivial angle $\theta$ is in turn the $T$--dual
of the magnetic flux). On the other hand, via a $S$--duality
transformation, one may obtain the M--theory counterpart of the type
IIA system described above. In fact, in computing the statistical
entropy, the M--theory  picture is more convenient to deal
with. Indeed, in this case the effective low--energy field  theory
describing the physics of the intersection is, in a suitable limit, a
conformal 2--dimensional one, rather than conformal quantum mechanics,
as it would be in the type IIA setting. Even if the two pictures describe
two $S$--dual regimes, the corresponding entropy, which is a
$U$--duality invariant, is the same and  so is the degeneracy of the
corresponding microscopic degrees of freedom. 

The M--theory configuration describing the generating solution is
depicted in table  1 and consists of a set of three bunches of
M5--branes intersecting on a  (compact) line along which some units of
KK--momentum has been put, and with  non--trivial 3--form potential
switched--on on their world volume (the latter  accounting for the
essential fifth parameter  characterizing the generating
solution). The compact  space $T_6\times S_1$ is along directions
$x^4,...,x^9,x^{10}$ while the  non--compact four dimensional
coordinates are $x^0,x^1,x^2,x^3$.  The M5--branes are
$N_1,N_2,N_3\,q^2$ respectively, there are $N_0+N_3\,p^2$ units of KK
momentum  along the spatial $10^{th}$ direction and the magnetic flux,
related to non--trivial 3--form field strength $h_{(3)}$ excited on
the M5--brane, is proportional to a rational number $\gamma=p/q$,
($p,q\in \ZZ$). The integers $p,q$ are related to the angle $\theta$
characterizing the $S\times T$--dual type IIB D3--brane configuration
by the condition: $q\, {\rm sin}\,\theta=p\, {\rm cos}\,\theta$.
\begin{table} [ht]
\begin{center}
\begin{tabular}{|c|c|c|c|c|c|c|c|c|c|c|c|}
\hline  Brane & 0 & 1 & 2 & 3 & 4 & 5 & 6 & 7 & 8 & 9 & 10 \\ \hline
$P_L$     & $\cdot$  &$\cdot$ &$\cdot$ &$\cdot$ &$\cdot$ &$\cdot$
&$\cdot$ &$\cdot$ &$\cdot$ &$\cdot$ & $\times$ \\   \hline $M5$ &
$\times$ &$\cdot$ &$\cdot$ &$\cdot$ &$\cdot$ &$\cdot$ &$\times$
&$\times$ &$\times$  &$\times$ & $\times$  \\   \hline  $M5$ &
$\times$ &$\cdot$ &$\cdot$ &$\cdot$ &$\times$ &$\times$ &$\cdot$
&$\cdot$ &$\times$  &$\times$ & $\times$  \\   \hline $M5+h_{(3)}$ &
$\times$ &$\cdot$ &$\cdot$ &$\cdot$ &$\times$ &$\times$ &$\times$
&$\times$ &$\cdot$   &$\cdot$ & $\times$  \\ \hline
\end{tabular}
\caption{\small The M--theory configuration corresponding to the
$N=8$,$d=4$ BPS black holes  generating solution.  The orientation
along different WV directions (which  has not been made explicit in
this table) should be the suitable one so to preserve
supersymmetry. We refer to \cite{gen} for details.}
\end{center}
\label{Mbrane}
\end{table}
As illustrated in \cite{gen}, the solution at the asymptotically flat
radial infinity is characterized by a set of eight moduli--dependent
charges $(y,x)\equiv (y^\alpha,x_\alpha),\,\,\alpha=0,\dots,3$ (four
magnetic $y^\alpha$ describing the Kaluza--Klein monopole  and M5
charges, and four electric $x_\alpha$ related to the Kaluza--Klein
momentum and M2 charges)\footnote{These charges are related to the
moduli--independent quantized charges through a symplectic
transformation (see \cite{mm} and \cite{gen}) and therefore are
quantized as well.}. The precise relation between the macroscopic
charges $(y,x)$ as related to the effective charges along the various
cycles  of $T_6\times S_1$ and the microscopic parameters
$\{N_\alpha,p,q\}$ in the M--theory description of our solution (as
well as in the $S$--dual type IIA one), is given in table
\ref{Mbrane2}. This correspondence was worked out in \cite{gen} and
was made possible thanks to an intrinsic group theoretical
characterization,  carried out in \cite{mm}, of the ten dimensional
origin of the vector  and scalar fields in the four dimensional $N=8$
model, once the  latter is interpreted as the low energy effective
theory of type  IIA/IIB superstring theories on $T_6$.
\begin{table} [ht]
\begin{center}
\begin{tabular}{|c|c|c|c|}
\hline  M-brane cycles & Type IIA cycles & Charges & 4D Charges  \\
\hline $KK$--monopole & $D6$(456789) & $0$ & $y_0$\\    \hline
$M5$(6789$|$10)& $D4$(6789) & $N_1$ & $y_1$ \\    \hline
$M5$(4589$|$10)& $D4$(4589) & $N_2$ & $y_2$ \\  \hline
$M5$(4567$|$10)& $D4$(4567) & $N_3 q^2$ & $y_3$ \\    \hline
$KK$--momentum &$D0$ & $N_0 + p^2 \; N_3$ & $x_0$ \\    \hline
effective $M2$(45)&effective $D2$(45) & $- p\, q \;N_3$ & $x_1$ \\
\hline effective $M2$(67)&effective $D2$(67) & $p\, q\;N_3$ & $x_2$ \\
\hline effective $M2$(89)&effective $D2$(89) & $0$ & $x_3$ \\    \hline
\end{tabular}
\end{center}
\caption{\small The effective charges along different cycles  of
$S_1\times T_6$ of the generating solution in terms of microscopic
parameters. The signs of the different charges are the correct ones so
to have a 1/8 susy preserving state. The last column gives the
effective charges in terms of moduli dependent quantized charges
$(y^\alpha,x_\alpha)$ defined in the supergravity framework and
characterizing the macroscopic description of the solution. They
depend on the suitably chosen asymptotic values of the scalars fields
at radial infinity, see \cite{gen} for details.}
\label{Mbrane2}
\end{table}

The expression of the entropy in terms of the charges $(y,x)$ is given
by applying to our solution \cite{gen} (for which $y^0=x_3=0$) the
Bekenstein--Hawking formula and expressing the area of the horizon
$A_h$ (in suitable units) in terms of the charges at infinity:
\begin{eqnarray}
\label{Smacro}
S &=& \frac{A_h}{4 l_P^2}=2\pi \sqrt{y_1y_2y_3 \left[x_0-\frac{(x_1
y_1-x_2 y_2)^2}{4y_1y_2y_3} \right]}
\end{eqnarray}
$l_P$ being the Plank length. The moduli dependence of the charges
$(y,x)$ in the above expression drops out, consistently with the
moduli independence of the entropy. Using table \ref{Mbrane2}, the
macroscopic entropy can be expressed in terms of the microscopic
parameters and reads: \beqa 
\label{entrmi} 
S = 2\,\pi\,\sqrt{N_1N_2N_3\, q^2\left[N_0 + p^2\, N_3- \frac{1}{4}
p^2\, N_3\frac{\left(N_1+N_2\right)^2}{N_1N_2}\right]} 
\label{Smicro}
\eeqa  
where the first two terms in 
the square bracket correspond to $x_0$ (the total momentum along
$S_1$) while the third term represents a shift $\Delta x_0$ to be
subtracted to $x_0$ in order to  define the relevant momentum
contribution to the entropy and which we shall comment on in the
sequel. Notice that the second and third terms  in the square bracket
are related to  non--trivial membrane effective charges and therefore
consistently vanish as  $\gamma,\, p\rightarrow 0$ (while $q$ can be
absorbed in a re--definition of $N_3$), i.e. when the magnetic flux,
in  the type IIA/M--theory language, or the non--trivial angle
$\theta$,  in the type IIB description, vanishes.

We wish to derive the expression (\ref{Smicro}) from a microscopic BPS
state counting.  As anticipated, this can be done along the lines of
\cite{msw,v1}. Having expressed the entropy in terms of charges
computed in the asymptotically flat radial infinity allows us to
perform  a ``far from the horizon'' counting of states for which the
relevant  framework is M--theory on $M_4\times T_7$ (or type II
superstring on  $M_4\times T_6$).  Although the analysis of
\cite{msw,v1} is concerned with BPS black holes deriving from
M--theory compactified on a manifold of the form $M_4\times CY_3\times
S_1$ which has a different topology, we expect the low energy
properties  of our solution at tree level to coincide with those of
the black holes  studied in  \cite{msw,v1} for a suitable choice of
the $CY_3$ manifold.  Indeed (at tree level) the generating solution
of $N=8$ regular BPS black  holes is also a solution of an $N=2$
consistent truncation of the $N=8$  model, namely the STU model, whose
six dimensional scalar manifold has a  geometry defined by a cubic
prepotential.  This model describes also the tree level low energy dynamics 
of black  holes within M--theory on $M_4\times CY_3\times
S_1$, where the prepotential  characterizing the special K\"ahler
geometry of the complex structure moduli space of $CY_3$ is cubic, at
tree level. We shall check, in this  particular case, that the result
of the entropy counting attained  in \cite{msw,v1} does coincide  at
tree level with the result of the analogous calculation we shall
perform on our generating solution. The following analysis will be
carried  out in the eleven dimensional framework, adapting to the
torus the study  in \cite{msw,v1}. References to the dual type IIA
setting will be done  under the reasonable assumption that the
degeneracy of BPS microstates  is insensitive to the type
IIA/M--theory duality\footnote{This would  require in turn that, in
changing the moduli of the background, singularities  are not
encountered and all the quantities defining BPS states behave
smoothly, which is indeed the case.}.

Let us first briefly comment on the regime of parameters in which our
computations are carried out. First of all we require the M--theory
modes  to decouple from the Kaluza--Klein modes of eleven dimensional
supergravity  compactified on $M_4\times T_7$\footnote{Which amounts
to asking  Kaluza--Klein supergravity to provide a reliable
description of the physics  on the chosen eleven dimensional
vacuum.}. This happens if $T_7$ is {\it  large} in the eleven
dimensional  Planck units, i.e. if $R_i\gg l_{P}$,  $R_i$ being  the
radii of the internal directions of $T_7$. Secondly we  demand the
four dimensional supergravity description of the generating  solution
to be reliable. This is the case if the  curvature of the solution
(whose upper bound is the near horizon curvature $\approx 1/A_h$) is
smaller than the scale fixed by the Kaluza--Klein spectrum. A
sufficient  condition for this to hold is therefore $A_h \gg R_i^2$.
Using eq.(\ref{Smacro}) the latter amounts to the  condition that the
quantized charges $(x,y)$ be much larger than $R_i/l_P$ ($\gg 1$).

A successful strategy for achieving a microscopic entropy counting has
been  so far to choose suitable limits in the background geometry such
that the  low energy quantum fluctuations around the solution are
described  by a two--dimensional conformal field theory on a torus  (a
cylinder in the limit of non--compact time). In this case the
asymptotic value of the degeneracy of microstates $\rho(h)$  for high
excited levels $h$ is given in terms of the central charges  of the
$\sigma$--model by the Cardy formula \cite{cardy,carlip}. The
conformal field theory considered  in \cite{msw} emerges in the limit
in which the radius of the eleventh dimension $R$ is much larger than
the radii of the remaining compact manifold, which means, in our case, 
$R\gg V(T_6)^{1/6}$. In this limit, the low energy fluctuations of the
three M5--branes will be independent of the $T_6$ coordinate and
described  by a conformal $1+1$ $\sigma $--model. On the six--dimensional 
world volume of each M5--brane embedded in the background of the other 2 branes
there is $(0,2)$ supersymmetry and BPS excitations will break half of
them. This chiral  supersymmetry will manifest on the two--dimensional
conformal field theory  as a $(0,4)$ supersymmetry (this is the same
effective theory describing  the microscopic degrees of freedom of the
configuration described in  \cite{msw,v1}). The BPS states  of this model will be annihilated by the
right--moving supercharges and  therefore they are described as
excitations of the left moving sector.  Denoting by $c_{\small L}$ the
left--mover central charge and by $h$ the  excitation level, the
asymptotic value of $\rho(h)$ for large $h$  ($h\gg c_{\small L}$) is
given by the Cardy formula\footnote{An other  condition for the
validity of this formula, which we can reasonably  assume to hold in
our case, is that the minimum excitation level $h_0$  should be small:
$h_0\ll c_{\small L}$.}:
\begin{equation}
\label{cardy} 
\rho(h) \approx  \left(\frac{c_{\small L}}{96 h^3}\right)^{1/4}
e^{2\pi \sqrt{\frac{ c_{\small L} h}{6}}}\rho(h_0) \approx e^{2\pi
\sqrt{\frac{ c_{\small L} h}{6}}}
\end{equation}
From the above expression, using the Boltzmann formula, one can derive
the  asymptotic value of the black hole entropy: 
\begin{equation}
\label{caren} S_{micro}=\ln\, \rho(h) \approx
2\pi\sqrt{\frac{c_{\small L}\, h}{6}} 
\end{equation} 
In the M--theory picture $h$ is the non--zero mode contribution to the
momentum along the eleventh direction $S_1$, i.e. in the conformal
theory language, the non--zero mode contribution to $L_0-\bar{L}_0$.
In the type IIA description the momentum along $S_1$ is given by the
D0--brane charge: $x_0=N_0+p^2\, N_3$. Clearly the regimes of validity
of the type IIA picture and the description in terms of the above
defined  conformal field theory are opposite. Since, as previously
said, the charge  and density of BPS states can be supposed to be
invariant with respect to  the $S$--duality mapping large to small
radius of $S_1$, we shall compute  $h$, in the type IIA framework, in
terms of a suitable contribution  to the D0--brane
charge. This is done in section 3.

\section{Computation of the central charge $c_L$}

Let us start by computing the central charge  $c_{\small L}$. 
The general expression for $c_{\small L}$ is:
\begin{equation} 
c_{\small L}=N_L^B+\frac{1}{2}N_L^F \label{cLdef} 
\end{equation} 
$N_L^B$ and $N_L^F$ being the number of left--moving bosonic and
fermionic degrees of freedom, respectively.  The number of
left--moving bosons has essentially two contributions. One is coming
from the moduli of the 4--cycles $P$ of  the torus $T_6$ along  which
the M5--branes are wrapped (call their number $d_p$). The other
contribution comes from the moduli of the rank two  antisymmetric
tensor potential $b$ propagating on the M5--branes world volume
($h_{(3)}=db$).  We can assume that each couple of M5--branes does 
intersect along the common directions (e.g. the branes with charges
$y_1$ and $y_3$ along the plane $(6,7)$)\footnote{An alternative is
for the projections of the  two M5--branes  along the common
directions to be separated by a distance along  the non--compact space
directions. The two configurations clearly have the  same charges and
energy.}. In this case the configuration of the three  M5--branes can
be described by a superposition $P$ of 4--cycles of $T_6$  whose {\it
fundamental class} is  $[P]= \sum_i\, y^i\, \alpha_i$ with
$\alpha_i,\,i=1,2,3$ being three $(1,1)$--cycles in $H^2(T_6,\ZZ)$
which will be defined precisely  for our configuration in the
sequel\footnote{We take the complex structure  on $T_6$ to be  defined
in this way: $z^1=x^4+{\rm i}\, x^5$,  $z^2=x^6+{\rm i}\, x^7$,
$z^3=x^8+{\rm i}\, x^9$.}  (recall that 2--cycles are isomorphic to 
4--cocycles).  As far as the first type of moduli is concerned one can
see, assuming  suitable conditions on $[P]$ \footnote{Along the lines
of  \cite{msw},  we assume the 4--cycle $P$ to be a {\it very ample}
divisor of  $T_6$, which formalizes in the language of algebraic
geometry the requirement for $P$ to be ``large'', besides the weaker
condition for $[P]$ to define a K\"ahler class (positiveness),
\cite{griff}.} and recalling that a torus  has vanishing Chern
classes, that:
\begin{equation}
d_p=\frac{1}{3}\int_{T_6} [P]^3 -2
\end{equation}
Let us now consider the contribution from the 2--form $b$. As
previously stated, in the limit of small $T_6$ with respect to  $S_1$,
the field $b$ can be considered, in the low energy limit, as  function
of just the coordinates $\{x^{10},x^0 \}$ on $S_1\times \IR$.  The
three form $h_{(3)}=d\,b$ defined on the M5 world volume is
self--dual.  This implies that among the fields $b$ which are 2--forms
on $P$, the left--moving ones are anti--self--dual ($b^-$), while the
right--moving  ones are self--dual ($b^+$). Moreover, the forms $b$
with just one index  on $P$ are non--dynamical gauge fields in the
theory on $S_1\times \IR$  which will enter the game since
$b_1=2h_{(1,0)}(P)\neq 0$ and on which  we are going to comment in a
moment. Using the Hodge index theorem on $P$  and doing some simple
calculations, one can see that the spaces of the $b^\pm$  have the
following dimensions, respectively: 
\begin{eqnarray}
{\rm dim}\{b^-\} &=& h_{(1,1)} - 1 = \frac{2}{3}\int_{T_6} [P]^3 + 2
h_{(1,0)} -1 \nn \\  {\rm dim}\{b^+\} &=& 1 + 2 h_{(2,0)} =
\frac{1}{3}\int_{T_6} [P]^3 +  2 h_{(1,0)} -1
\label{b+b-}
\end{eqnarray}
As far as the fermion modes $N^F_{L,R}$ are concerned, from standard
analysis it  is known that on a complex manifold the number of
left--moving and  right--moving fermions are related to the dimension
of $H^{(2r+1,0)}(P)$  and $H^{(2r,0)}(P)$, respectively:
$N_L^F=4h_{(1,0)}$ and $N_R^F=4(h_{(0,0)}+h_{(2,0)})$. Let us now
collect our results.  With the above definitions, the number of left
and right moving bosons is given, in general, by:
\begin{eqnarray}
N_L^B &=& d_p + {\rm dim}\{b^-\} + 3 \nn \\ N_R^B &=& d_p + {\rm
dim}\{b^+\} + 3
\end{eqnarray} 
where the $+3$ in the two cases takes into account the contribution of
three translational zero modes. Hence we have:
\begin{eqnarray}
\label{bf}
N_L^B&=&\int_{T_6} [P]^3 + 2 h_{(1,0)} \,\,\,\{- 2 h_{(1,0)} =
\int_{T_6} [P]^3\} \nn \\ N_L^F&=&4 h_{(1,0)} \,\,\,\{- 4 h_{(1,0)} =
0\} \nn \\  N_R^B&=&\frac{2}{3}\int_{T_6} [P]^3 + 2 h_{(1,0)}
\,\,\,\{- 2 h_{(1,0)} = \frac{2}{3}\int_{T_6} [P]^3\}  \nn \\
N_R^F&=&4 h_{(2,0)} + 4 \,\,\,\{- 4 h_{(1,0)}\}
\end{eqnarray}
where the terms in the curly brackets represent the effect of the
gauging  of the $b_1$ non--dynamical gauge fields introduced
previously. Indeed, coupling  the (left and right moving) bosonic and
fermionic modes to these vector fields  will reduce the scalar degrees
of freedom by $b_1=2h_{(1,0)}(P)$ and the fermionic ones by $2\, b_1$
(we require, according to \cite{moooa}, this coupling to be
left--right symmetric). This gauging is {\it not} optional, since it
restores supersymmetry on the  right sector. This can be easily seen
deriving from eq.(\ref{b+b-}) the following relation:
\begin{equation}
\frac{1}{6}\int_{T_6} [P]^3 = h_{(2,0)} - h_{(1,0)} + 1
\end{equation}
which in turn implies that, provided the gauging is performed, the 
values of $N_R^B$ and $N_R^F$, read off from eq.s (\ref{bf}), coincide. 
From eq.(\ref{cLdef}) and eq.s (\ref{bf}) we can finally
deduce the value of the  central charge $c_{\small L}$ to be:
$c_{\small L}=\int_{T_6}[P]^3$.  To compute its value in terms of the
quantized charges $y^i$ we consider  suitable representatives of the
classes $\alpha_i$ and define as $D_{ijk}$ the restriction to them of the triple 
intersection numbers of $T_6$:  
\begin{eqnarray}
1 &=& 6 D_{ijk}=\int \alpha_i\wedge\alpha_j \wedge\alpha_k\nonumber\\
\alpha_i & =& dx^a\wedge dx^b\nonumber\\ \{i\} &=& \{1,2,3\}\equiv
\{(ab)\}=\{(45),(67),(89)\}
\label{iab}
\end{eqnarray}
The left--mover central charge is therefore:
\begin{equation}
c_{\small L}=\int_{T_6} [P]^3 = 6\, y_1\, y_2\, y_3=6\, q^2\,N_1\,
N_2\, N_3
\label{cLmicro}
\end{equation}
Notice that the gauging discussed earlier besides ensuring
supersymmetry  on the right--mover sector, provides a consistency
condition for the left  sector as well. Indeed if it were not
performed, $c_{\small L}$ would have an additional term proportional
to $b_1$. As a consequence of this, the  entropy computed from
eq.(\ref{caren}), provided $h\neq 0$, would have a non  vanishing
leading contribution even if the number of intersecting (bunches  of)
branes would have been less than three\footnote{ As explained above we are considering the limit
where $N_0>>c_{\small L}$. In this regime, the dominant configuration
is that  of ``short'' branes \cite{ms}: we have three bunches of
parallel M5--branes, $N_1,N_2,N_3\,q^2$ respectively which therefore
intersect on  $N_1\,N_2\,N_3\,q^2$ points on $T_6$.}, in disagreement with the
Bekenstein--Hawking  macroscopic prediction (differently, in the case
of a generic C--Y manifold one needs instead just one set of coinciding
branes to have a regular horizon in four dimensions, see for instance \cite{bfis}). 

\section{Non--zero mode contribution to the momentum along $S_1$: a type IIA computation}

The remaining quantity to be computed on our
solution is the excitation  level $h$ in eq.(\ref{caren}).  This can
be done by extending to our configuration on $T_6$ the computation
performed in \cite{msw} within the M--theory framework. As previously
anticipated, we shall adopt the type IIA viewpoint instead, and write $h$ as the
difference between the {\it total} D0--brane charge $x_0=N_0+p^2\, N_3$
related to the background configuration and a contribution $\Delta
x_0$ to be suitably  interpreted from type IIA perspective. While the
former quantity corresponds in eleven dimensions to the total KK momentum along  
$x^{10}$, i.e., in the CFT
limit, to the eigenvalue of $L_0-\bar{L}_0$, the latter defines the zero--mode 
contribution to $L_0-\bar{L}_0$. On the type IIA side $\Delta x_0$, as we shall see, 
can be expressed in terms of the contribution to the D0--brane charge
due to a magnetic flux ${\cal F}^{(0)}$, defined on the intersections
along the planes $(45)$, $(67)$, $(89)$ of the D4--branes, which can be
interpreted as the {\it part} of the total magnetic flux ${\cal F}$
due to the same modes or states which in the $S$--dual CFT picture are
zero--modes of the potential $b$. In the language of open strings
attached to D4--branes these states can be possibly identified with
modes of  Dirichelet--Dirichelet strings  connecting the $N_3\, q^2$
branes to the $N_1$ and $N_2$ branes, which  are massless and will
induce an equal flux density ${\cal F}^{(0)}$ on the  world volume of
the $N_i$ and $N_j$ branes along their common directions. 
The flux ${\cal F}^{(0)}$ will correspond, upon
dimensional oxidation and suitable dualization, to a 3--form
$h_{(3)}^{(0)}$ on the M5--branes. The derivation of the WZ term in
the world volume action of the D4--brane, yielding the contribution of
the magnetic flux to the D0--brane charge, from the corresponding term
in the M5--brane world volume action, is discussed in the appendix, and
the result can be applied to the fields ${\cal F}^{(0)}$ and $h_{(3)}^{(0)}$. 

The magnetic flux density ${\cal F}^{(0)}$, differently from the whole
${\cal F}$  which is non--vanishing only on the world volume of the
$N_3\,q^2$ branes, can be defined on $T_6$ in terms of the three
2--forms $\alpha_i$:  ${\cal F}^{(0)}=\sum_i \,{\cal F}^{(0)|i}\,
\alpha_i$\footnote{The components $\{{\cal F}^{(0)|1},\,{\cal
F}^{(0)|2},\,{\cal F}^{(0)|3}\} $ correspond to $\{{\cal
F}^{(0)}_{45},\,{\cal F}^{(0)}_{67},\,{\cal F}^{(0)}_{89}\}$,
consistently with the  convention on the indices defined in
eq.(\ref{iab}).}. Its value is determined  in terms of the effective
electric charges $x_i$ through the WZ terms which  couple it to the D4
brane world volumes. The part of the R--R 3--form $A$ coupled to
${\cal F}^{(0)}$ in these terms is  $\sum_i A_\mu^i\,dx^\mu\wedge
\alpha_i$, where the index $\mu$ runs along  the non--compact
directions and the vectors fields $ A_\mu^i$ denote three  electric
potentials of the effective four dimensional theory. The corresponding
charges $x_i$ are defined by the four dimensional minimal  couplings:
\begin{eqnarray}
x_i\int\, A^i_0\, dx^0 = \frac{1}{2\pi}\,\sum_k\,\int_{(D4)_k} \,{\cal
F}^{(0)|j}\, \alpha_j\wedge \alpha_i\wedge A^i_0\, dx^0= \left(6\,
D_{ijk}\,y^k\,\, \frac{{\cal F}^{(0)|j}}{2\pi}\right)\,\times\, \int
A^i_0 \, dx^0 \nn \\
\label{gi}
\end{eqnarray} 
hence:   
\beqa  
x_i &=& 6\, D_{ijk}\,y^j\, G^{k}=6\, D_{ij}\, G^{j}
\label{gi2}
\eeqa  
where we have defined $G^i=F^{(0)|i}/(2\, \pi)$ and
$D_{ij}=D_{ijk}\,  y^k=\int_P \,\alpha_i\wedge\alpha_j/6$.  The
integrals on the right hand side of eq.(\ref{gi}) are computed on
the D4--brane world volumes, and eventually extended to the whole $T_6$ 
using the fundamental class $[P]$ previously
defined.  We have considered moreover ${\cal F}^{(0)|i}$ to be uniform
along the planes on  which it is defined.  The contribution of ${\cal
F}^{(0)}$ to the total D0--brane charge is given by:
\begin{eqnarray}
\Delta x_0 &=&-\frac{1}{2\,(2\pi)^2}\, \int_{P} {\cal F}^{(0)}\wedge
{\cal F}^{(0)}=3\, D_{ij}\,G^i\,G^j
\label{dx01}
\end{eqnarray} 
Inverting eq.(\ref{gi2}) one finds $G^i=D^{ij}\,x_j/6$,
where $D^{ij}\,D_{jk}=\delta^i_k$. Inserting this result in
eq.(\ref{dx01})  we obtain:
\begin{eqnarray}
\Delta x_0 &=&\frac{1}{4\,y_1\,y_2\,y_3}\left(\sum_i\,(y_i\,x_i)^2
-2\sum_{i<j}y_i\,x_i\, y_j\,x_j\right)= \frac{p^2\, N_3}{4\, N_1\,
N_2}\left(N_1+N_2\right)^2
\end{eqnarray}
where in the last passage we have used the expression of the charges
$(x,y)$ given in table \ref{Mbrane2}. This quantity,
being related in the $S$--dual CFT picture, to the zero--mode
contribution  to the KK momentum along $S_1$, has to be subtracted
from  the total D0--brane charge $x_0$ in order to obtain $h$. Hence
we finally get:
\begin{equation}
h= x_0-\Delta x_0=N_0+p^2\, N_3 -\frac{p^2\,
N_3}{4\,N_1\,N_2}\left(N_1+N_2\right)^2
\end{equation}
Inserting the above expression and eq.(\ref{cLmicro}) in
eq.(\ref{caren}) one finally obtains the correct microscopic
prediction for the Bekenstein--Hawking entropy, eq.(\ref{entrmi})!
This ends the computation.

\section{Conclusion}
Our final formula is the same as (3.28) of \cite{msw}. However we wish
to emphasize that the above computed microscopic entropy formula
accounts for the  entropy of {\it all} regular BPS black holes within
toroidally compactified string (or M) theory, \cite{gen}. In
particular,  the {\it shift} factor encodes the essential fifth
parameter which is crucial for the solution to be a generating
one. Notice that since the near horizon geometry is characterized by
just one of  the 5 parameters (i.e. the entropy or horizon area) which
is the one accounting for the microscopic degeneracy of the (proper)
black hole states\footnote{Here by ``proper'' black hole  we mean the
near horizon solution which is {\it hairless} thanks to supersymmetry,
as opposite to the whole solution interpolating between the horizon
and the asymptotically flat radial infinity, which has indeed {\it
hair}, i.e. it depends on matter fields.  The microscopic degrees of
freedom of this {\it hair} are encoded in four of the five
$U$--duality invariants.}, we actually expect  that this geometry does
not distinguish between regular solutions characterized by different
numbers of $U$--duality invariants. Indeed, for  suitable values of
the corresponding quantized charges, two different solutions can have
precisely the same near horizon geometry  (and therefore the same
entropy). For instance, the difference between the expression of  the
entropy for a five or a four parameter solution may amount just to the
aforementioned shift which can be absorbed in a re--definition of the
quantized charges (as far  as near horizon geometry is
concerned). However, if we wish to characterize the entropy of a
(proper) regular black hole as  part of the most general interpolating
solution, then the parameter characterizing this shift is an essential
ingredient for  supporting all the independent $U$--duality invariant
degrees of freedom. Therefore it is far from the horizon where this
shift  parameter acquires a highly non--trivial physical meaning. The
magnetic flux manifests itself by coupling to new scalars (the axions) 
whose radial evolution in the solution is non--trivial. These extra 
scalars are needed for the  solution to be a generating one, \cite{gen}.

\vskip 10pt
\noindent
{\bf Acknowledgements}

We would like to thank M. Bill\`o, G. Bonelli, R. Russo, C. Scrucca
and  A. Westerberg for  clarifying discussions and the Spinoza
Institute, where  part of this work has  been carried out, for the
kind ospitality. We acknowledge partial support by  ECC under
contracts  ERBFMRX-CT96-0045 and ERBFMRX-CT96-0012. M.B. is supported
by INFN.

\appendix
\section*{Appendix}

Let us briefly review the derivation of the WZ term describing the
contribution to the D0--brane charge due to a magnetic flux ${\cal F}$
on the D4--brane from the perspective of M--theory compactified on a
circle. The normalization we choose  for this term in the low energy
effective action on the D4--brane is the following:
\begin{equation}
-\,\frac{1}{2(2\pi)^2}\int_{P}{\cal F}\wedge {\cal F}\, \int C_{(1)}
\label{ambiguo}
\end{equation}
where $C_{(1)}$ is the pull--back of the R--R 1--form which  couples
to the D0--brane and $P$ is the four cycle of $T_6$ on which the
D4--brane is wrapped. 

In general, the derivation of the D4--brane action from that of the
M5--brane upon double dimensional reduction is not straightforward. In fact, as shown in detail in
\cite{sc1,sc2}, by compactifying the world volume action of the
M5--brane one ends in an dual gauge. In  order to get the usual
D4--brane action one has to perform a (electro--megnetic) 
duality  transformation,
which moreover mixes the DBI and the WZ terms. Here, however,  we are
just interested in deriving the term (\ref{ambiguo}),  and
computations can be much simplified. If we take the background gauge
potential to have only the  temporal component non--vanishing and
choose static gauge,  then only $C_0$ survives on the world volume
(and coincides with the corresponding  component of the background
field). If we now choose it to be small, the WZ term (\ref{ambiguo}) 
turns out to be the same in the two dual regimes, at leading 
order. This can be easily seen by comparison of formulae (81) and 
(82) of the appendix of \cite{sc2}.  

Let us then consider the theory on the M5--brane (double  oxidation of
the D4--brane) embedded in a Minkowsky space--time $G_{MN}=\eta_{MN}$
($M,\,N=0,\dots, 10$) and perform an infinitesimal shift on this
metric by a quantity $\delta\,G_{MN}$ whose only non zero entry is
$\delta\,G_{0\,10}=C_0\ll 1$ (we are thinking  of the dimensional
reduction on $S_1$ to ten dimensions and of the known  fact that
ultimately $C_m\equiv \delta\,G_{10\, m}$).  The world volume action
of the M5--brane would acquire a term of the form:
\begin{eqnarray}
\delta S &=& - \int_{S_1\times \IR} T_{vw}\delta\,\hat G^{vw}\;
d\xi^{10} \, d\xi^0\,=\, -2\int_{S_1\times \IR} T_{0\,10}\, C_0\;
d\xi^{10} \, d\xi^0 \nonumber\\  &&\nonumber\\ T_{0\, 10} &=&\beta\,
\left({h_{0}}^{a\,b}\,h_{10\,c\,d}\right)\,\int_{P}\alpha_{a\,b}\wedge
{}^*\alpha^{c\,d}\, \propto\, h_{0\,a\,b}\,{h_{10}}^{a\,b}
\label{pesce}
\end{eqnarray}
where it is understood that the M5--brane world volume extends over
$P\times S_1\times \IR$, the indexes $v,w$ get values $0,...,6,10$,
$\hat G$ is the pull--back of the metric $G$, and $*$ is the Hodge
duality on $P$. The constant $\beta$ is determined by our choice  of
normalization in eq.(\ref{ambiguo}) and will be fixed at the  end. If
we label by $a$ the internal directions of $T_6$ then we  choose the
components $h_{a\,b\,c}$ of the self--dual tensor to be zero. The
self--duality of $h_{(3)}$ implies that ${h_{\pm}}^{a\,
b}\,\alpha^\mp_{a\, b}=0$, where ``$\pm$'' labels on the $h_{(3)}$
tensor the light--cone coordinates $\xi^\pm=\xi^0\,\pm\,\xi^{10}$ and  on
the 2--form $\alpha$ its self--dual/anti--self--dual components in
$P$: $\alpha=\alpha^++\alpha^-$, ${}^*\alpha=\alpha^+-\alpha^-$.

We may rewrite the action term in eq.(\ref{pesce}) as follows:
\begin{eqnarray}
\delta S &=& - 2\beta\int_{P\times
S_1}\,{h_{0}}^{a\,b}\,h_{10\,c\,d}\,\alpha_{a\,b}\wedge
{}^*\alpha^{c\,d} \wedge d\xi^{10}\,  \int C_0\, d\xi^0 = \nonumber\\ &&-
2\beta\int_{P\times
S_1}\,\left({h_{+}}^{a\,b}\,h_{+\,c\,d}-{h_{-}}^{a\,b}\,h_{-\,c\,d}\right)
\,\alpha_{a\,b}\wedge {}^*\alpha^{c\,d} \wedge d\xi^{10}\, \int C_0\,
d\xi^0 = \nonumber\\ && - 2\beta\int_{P\times
S_1}\,{h_{10}}^{a\,b}\,{h_{10}}^{c\,d}\,\alpha_{a\,b}\wedge
\alpha_{c\,d} \wedge d\xi^{10}\, \int C_0\, d\xi^0
\label{fish}
\end{eqnarray}
where we have used the self--duality of $h_{(3)}$.

If we integrate the expression in eq.(\ref{fish}) over $S_1$ (very
small), after considering only the zero--modes along it, and make the
(leading--order) identification ${\cal F}_{a\, b}/(2\pi)=h_{10\,a\,b}$
(see \cite{howest}) we obtain precisely the term in eq.(\ref{ambiguo}) setting
$\beta=1/(8\,\pi\,R)$, $R$ being the radius of  the eleventh dimension.



\begin{thebibliography}{99}

\bibitem{gen} M. Bertolini and M. Trigiante, Nucl. Phys.  {\bf B582}
(2000) 393, hep-th/0002191.

\bibitem{mm} M. Bertolini and M. Trigiante, Int. J. Mod. Phys. {\bf
A} (in press), hep-th/9910237.

\bibitem{malda} J. Maldacena, {\it ``Black Holes in String Theory''}, hep-th/9607235. 

\bibitem{amanda} A.W. Peet, Class. Quant. Grav. {\bf 15} (1998) 3291,
hep-th/9712253.

\bibitem{mooa} T. Mohaupt, {\it ``Black Hole Entropy, Special Geometry
and Strings''},  hep-th/0007195.

\bibitem{bala} V. Balasubramanian, {\it ``How to Count the States of
Extremal Black Holes in $N=8$ Supergravity''}, hep-th/9712215.

\bibitem{msw} J. Maldacena, A. Strominger and E. Witten, JHEP {\bf 12}
(1997) 2, hep-th/9711053.

\bibitem{v1} C. Vafa, Adv. Theor. Math. Phys. {\bf 2} (1998) 207,
hep-th/9711067.

\bibitem{moooa}G.L. Cardoso, B. de Wit and T. Mohaupt,
Nucl. Phys. {\bf B567} (2000) 87, hep-th/9906094.

\bibitem{sen1} A. Sen, Mod. Phys. Lett. {\bf A10} (1995) 2081,
hep-th/9504147.

\bibitem{griff} P. Griffiths, J. Harris, {\it ``Principles of
Algebraic Geometry''}, Wiley--Interscience (1978).

\bibitem{cardy} J. Cardy, Nucl. Phys. {\bf B270} (1986) 186.

\bibitem{carlip} S. Carlip, {\it ``Logarithmic Corrections  to Black
Hole Entropy from  the Cardy Formula''}, gr-qc/0005017.


\bibitem{bfis} M. Bertolini, P. Fr\'e, R. Iengo and C.A. Scrucca, 
Phys. Lett. {\bf B431} (1998) 22, hep-th/9803096.

\bibitem{ms} J. Maldacena and L. Susskind, Nucl. Phys. {\bf B475}
(1996) 679, hep-th/9604042.

\bibitem{sc1} M. Aganagic, J. Park, C. Popescu and J.H. Schwarz,
Nucl. Phys. {\bf B496}  (1997) 191, hep-th/9701166.

\bibitem{sc2} M. Aganagic, J. Park, C. Popescu and J.H. Schwarz,
Nucl. Phys. {\bf B496}  (1997) 215, hep-th/9702133.

\bibitem{howest} P.~S.~Howe, E.~Sezgin and P.~C.~West,
Phys. Lett. {\bf B399} (1997)  49,  hep-th/9702008.

\end{thebibliography}
\end{document}